\documentclass[aps,prx,onecolumn,nofootinbib,longbibliography]{revtex4-2}
\usepackage{amsmath,amssymb,amsfonts,bm}
\IfFileExists{physics.sty}{\usepackage{physics}}{}
\providecommand{\ket}[1]{\lvert#1\rangle}
\usepackage{graphicx}
\usepackage{hyperref}
\usepackage{mathtools}
\usepackage{mathtools}
\usepackage{amscd}
\usepackage{amsfonts}
\usepackage{amsmath}
\usepackage{amssymb}
\usepackage{mathrsfs}
\usepackage{color}
\usepackage{bm}
\usepackage{graphicx}
\usepackage{graphics}
\usepackage{wrapfig}
\usepackage[utf8]{inputenc}
\usepackage{gensymb}
\usepackage{booktabs}
\usepackage{subcaption}
\usepackage[margin=1in]{geometry}
\usepackage{tikz}
\usepackage{ragged2e}
\usetikzlibrary{decorations.pathmorphing}
\def\>{\rangle}
\def\<{\langle}

\def\sss{\scriptscriptstyle}

\begin{document}

\title{Spectral Softening and the Structural Breakdown of Thermodynamic Equilibrium}

\author{Ilki Kim}
\email{hannibal.ikim@gmail.com} \affiliation{Department of Physics,
North Carolina A$\&$T State University, Greensboro, NC 27411}

\date{\today}

\begin{abstract}
Reversible quasistatic thermodynamics requires normalizable
instantaneous equilibrium states and dynamics allowing their
attainment or tracking. We examine these requirements in a driven
two-dimensional quadratic Hamiltonian with an angular-momentum
coupling and a spatially uniform momentum shift $p_c(t)$,
representable by a time-dependent vector potential. Its undriven
part has normal-mode frequencies $\omega_\pm=\omega_1\pm\omega_2$
and the Fock--Darwin structure of a charged oscillator in a
perpendicular magnetic field. At the exact soft-mode endpoint
$\omega_-=0$, confinement is lost in one normal-mode sector. For
every strictly positive $\omega_-$, the canonical Gibbs state
remains normalizable; only at the endpoint does the canonical trace
cease to be finite. In the quasi-soft regime, where
$\omega_-\to0^+$, exact adjacent-level couplings yield a
dimensionless adiabatic parameter proportional to $|\dot
p_c|/\omega_-^2$. The standard adiabatic criterion requires this
parameter to remain much smaller than unity. At a fixed occupation
and a fixed nonzero driving rate, however small, this parameter
grows without bound as $\omega_-\to0^+$, eventually violating this
criterion. In an open system, thermal tracking additionally depends
on relaxation relative to the driving timescale. An explicit linear
system--bath coupling illustrates why the relaxation rate cannot be
inferred from the soft-mode spacing alone. Finally, the thermal
Wigner function and its classical limit show that phase-space
confinement is lost in both descriptions at the soft-mode endpoint.
Taken together, our results show that canonical equilibrium is lost
at $\omega_-=0$. For $\omega_->0$, the Gibbs state exists, while the
ability to track it dynamically depends on the driving and
dissipation.
\end{abstract}

\maketitle

\section{Introduction}
%
Quasistatic thermodynamics describes reversible evolution through a
sequence of equilibrium states, provided that equilibrium can be
established and maintained throughout the process
\cite{LandauStatMech}. This description requires both the existence
of a normalizable equilibrium state for the instantaneous
Hamiltonian and dynamical conditions that allow the system to attain
or approximately track that state. For an isolated quantum system,
slow parameter variation can support adiabatic following of
instantaneous eigenstates
\cite{BornFock,Kato,Berry,Torrontegui2013}, a process distinct from
thermal equilibration. For a system in contact with a thermal
environment, the ability to attain or track an instantaneous Gibbs
state depends on the system--bath coupling and the associated
relaxation timescales.

Spectral gaps, gap closing, and level crossings have long been
central to analyses of quantum adiabaticity and nonadiabatic
transitions
\cite{BornFock,Kato,PolkovnikovRMP2011,AlbashLidar2018,WignerNeumann1929,Landau1932,Zener1932,Shevchenko2010};
see also Ref.~\cite{Berry}. In these analyses, the spectral
structure and the driving protocol jointly determine the conditions
for adiabatic following and the emergence of nonadiabatic
transitions. Related questions arise in critical phenomena, where
soft modes, whose characteristic frequencies approach zero, play a
central role in critical slowing down and the divergence of
relaxation times
\cite{HohenbergHalperin1977,FolkMoser2006,Tauber2014,Goldenfeld1992}.
Such divergent relaxation times can coexist with a well-defined
equilibrium description, even at criticality. The slowing of
relaxation must therefore be distinguished from the loss of a
normalizable equilibrium state. These considerations raise the
question of how spectral softening affects both the existence of
equilibrium and its dynamical accessibility.

Here we address these questions in a minimal two-dimensional
quadratic Hamiltonian comprising an isotropic oscillator of
frequency $\omega_1$, an angular-momentum coupling $\omega_2\hat
L_z$, and a time-dependent, spatially uniform momentum shift
$p_c(t)$. Its undriven part has two normal modes with frequencies
$\omega_\pm=\omega_1\pm\omega_2$. In addition to admitting a
rotating-frame description in terms of two uncoupled harmonic
oscillators of frequency $\omega_1$, this undriven Hamiltonian can
be physically realized as a harmonically confined charged particle
in a perpendicular magnetic field, as described by the Fock--Darwin
model \cite{Fock1928,Darwin1931,ReimannManninen2002}; see also
Ref.~\cite{GriffithsSchroeter2018} for a textbook discussion of
Landau levels. We approach the soft-mode endpoint from the strictly
confining regime $\omega_1>\omega_2>0$. The soft mode thus emerges
through competition between two finite frequency scales, with
neither $\omega_1$ nor $\omega_2$ vanishing. At $\omega_1=\omega_2$,
the rotating-frame resonance coincides with the disappearance of a
confining normal-mode scale in the laboratory-frame Hamiltonian.

Our analysis yields two distinct results. The first concerns the
\emph{existence} of instantaneous equilibrium. For every strictly
positive $\omega_-$, however small, the instantaneous Hamiltonian is
bounded below and has two positive normal-mode frequencies, and so a
normalizable Gibbs state exists at any finite positive temperature.
At fixed inverse temperature $\beta>0$, the soft-mode partition
factor grows as $Z_-\sim(\beta\hbar\omega_-)^{-1}$ as
$\omega_-\to0^+$, but remains finite at every positive $\omega_-$.
Throughout this finite-gap regime, the statistical weights therefore
remain well defined and normalized. This divergence as
$\omega_-\to0^+$ does not imply a loss of equilibrium at any finite
positive gap. The canonical state ceases to be normalizable only at
the exact endpoint $\omega_-=0$, where confinement is lost in the
soft-mode sector. This minimal quadratic model thus allows us to
examine the thermodynamic consequences of spectral softening without
invoking many-body complexity or critical phenomena.

In contrast, the second result concerns \emph{dynamical
attainability} within the finite-gap regime, where instantaneous
equilibrium remains well defined. For the specified momentum-shift
drive, the exact nonadiabatic couplings connect neighboring
normal-mode levels. As derived in Sec.~\ref{sec:nonadiabatic}, the
soft-mode raising channel yields a dimensionless adiabatic parameter
proportional to $|\dot p_c|/\omega_-^2$, with an explicit
occupation-number factor. The standard adiabatic criterion requires
this parameter to remain much smaller than unity. At a fixed
occupation and a fixed nonzero driving rate, however small, this
parameter grows without bound as $\omega_-\to0^+$. Consequently, the
small-parameter condition is no longer satisfied below a drive- and
occupation-dependent frequency scale, even though the gap remains
positive, the Hamiltonian remains bounded below, and the Gibbs state
remains normalizable. The calculation thus quantifies a
drive-dependent restriction on the standard adiabatic approximation
at finite positive gaps, where canonical equilibrium remains well
defined.

To assess the implications for quasistatic thermodynamic
reversibility, we next ask whether bath-induced relaxation can
sustain thermal tracking under the imposed driving as
$\omega_-\to0^+$. We consider an illustrative linear system--bath
coupling to the soft mode, which permits only adjacent-level
transitions at leading order. The resulting occupation-relaxation
time is $\tau_{\mathrm{occ}}=1/\kappa(\omega_-)$, where the damping
coefficient $\kappa(\omega_-)$ is determined by the specified
coupling and bath correlations. Thermal tracking in the quasi-soft
regime therefore requires comparing the relevant relaxation
timescales with the timescale on which the imposed driving changes
the instantaneous Gibbs state. Although this state remains
normalizable at every strictly positive $\omega_-$, tracking
maintained by the bath requires sufficiently rapid relaxation
relative to the imposed driving.

Finally, the Wigner phase-space representation
\cite{Wigner,Hillery,Zurek2001} and its classical limit reveal the
geometric origin of the endpoint singularity. For every strictly
positive $\omega_-$, the thermal phase-space distribution remains
normalizable, while it broadens as $\omega_-$ decreases. Only at the
exact endpoint $\omega_-=0$ is confinement lost in one normal mode,
while the other, with frequency $\omega_+$, remains confined. This
\emph{partial deconfinement} prevents a normalizable canonical
distribution in both the quantum and classical descriptions. The
quantum--classical correspondence thus shows that the endpoint loss
of equilibrium is not a purely quantum effect, but reflects the
geometry of Hamiltonian phase space.

Taken together, these results show that slow driving alone does not
guarantee quasistatic thermodynamic reversibility. Such a
description presupposes both a normalizable equilibrium measure and
dynamical conditions allowing the system to attain or track it. Our
central conclusion is the distinction between the
\emph{in-principle} existence of a canonical equilibrium state for
every $\omega_->0$ and its \emph{physical attainability} under the
specified driving and dissipation. Only at the exact soft-mode
endpoint, reached continuously from the confining regime, does the
normalizable equilibrium measure itself cease to exist. Reversible
quasistatic evolution through canonical equilibrium states is then
precluded by the loss of equilibrium itself.

The paper is organized as follows. Sec.~\ref{sec:model} defines the
model and clarifies the rotating-frame and magnetic-oscillator
interpretations. Sec.~\ref{sec:spectral_structure} derives the
instantaneous spectral structure. Sec.~\ref{sec:nonadiabatic}
quantifies the drive-dependent adiabatic criterion.
Sec.~\ref{sec:thermodynamic} separates bath-dependent thermal
relaxation from the exact partition-function singularity.
Sec.~\ref{sec:wigner} develops the Wigner and classical phase-space
descriptions. Sec.~\ref{sec:discussion} provides concluding remarks.

\section{Model}\label{sec:model}
%
We consider a driven quadratic Hamiltonian in two dimensions,
\begin{equation}
    \hat{H}_1(t) = \frac{(\hat{p}_x + p_c(t))^2}{2m}
    + \frac{(\hat{p}_y + p_c(t))^2}{2m}
    + \frac{m\omega_1^2}{2}(\hat{x}^2 + \hat{y}^2)
    + \omega_2 \hat L_z\,,\label{eq:H1}
\end{equation}
where the orbital angular momentum is $\hat L_z = \hat x \hat p_y -
\hat y \hat p_x$. The time dependence enters solely through the
$c$-number momentum shift $p_c(t)$, which may be arbitrary. This
momentum shift in the kinetic terms can be represented by a
spatially uniform vector potential with $A_x(t)=A_y(t)=A(t)$,
through $p_c(t)=-(q/c)A(t)$. For detailed discussions of the
vector-potential description of spatially uniform, time-dependent
electric fields, see, e.g.,
Refs.~\cite{KriegerIafrate1986,IafrateSokolovKrieger2017}.

For $p_c=0$, the Hamiltonian decomposes into two independent
oscillators,
\begin{equation}
    \hat H_0 =
    \hbar\omega_{\sss{+}}\!
    \left(\hat n_{\sss{+}} + \frac{1}{2}\right)
    + \hbar\omega_{\sss{-}}\!
    \left(\hat n_{\sss{-}} + \frac{1}{2}\right)
    \,,\label{eq:H0}
\end{equation}
with $\omega_{\sss{\pm}}=\omega_1\pm\omega_2$. In the strictly
confining regime $\omega_1>\omega_2>0$, the spectrum consists of
discrete levels $E^{(0)}_{n_{\sss{+}},n_{\sss{-}}}$ with eigenstates
$\ket{n_{\sss{+}},n_{\sss{-}}}_{\scriptstyle 0}$ (see
Appendix~\ref{sec:S1} for details). When $\omega_2>\omega_1$, the
coefficient of $n_-$ in Eq.~\eqref{eq:H0} is negative, so the energy
decreases without bound as $n_-$ increases at fixed $n_+$. This
regime admits no normalizable canonical Gibbs state
\cite{Ruelle1969,LiebSeiringer2010} and is excluded from our
analysis.

The undriven Hamiltonian can also be written as $\hat H_0=\hat
H_{\mathrm{iso}}+\omega_2\hat L_z$. A transformation to a frame
rotating at the appropriate angular velocity removes the
angular-momentum term, leaving $\hat H_{\mathrm{iso}}$, which
describes two uncoupled Cartesian oscillators of frequency
$\omega_1$. In the laboratory frame, $\hat H_0$ can be realized as
the Fock--Darwin Hamiltonian of a harmonically confined charged
particle in a perpendicular magnetic field. The canonical trace of
the undriven system is evaluated here for this laboratory-frame
Hamiltonian, whose lower normal-mode frequency
$\omega_-=\omega_1-\omega_2$ vanishes at $\omega_1=\omega_2$.

Our analysis concerns the driven laboratory-frame Hamiltonian in
Eq.~\eqref{eq:H1}, with the prescribed momentum-shift drive
$p_c(t)$. We examine the limit $\omega_-\to0^+$ from the strictly
confining regime $\omega_1>\omega_2>0$. At each fixed $p_c$, a
phase-space displacement determines the instantaneous eigenstates
and their common energy shift, as shown in
Sec.~\ref{sec:spectral_structure}. Section~\ref{sec:nonadiabatic}
then examines the nonadiabatic couplings induced by the drive, and
Sec.~\ref{sec:thermodynamic} discusses the role of bath-induced
relaxation in thermal tracking.

\section{Spectral Structure Near the Soft-Mode
Limit}\label{sec:spectral_structure}
%
In the quasi-soft regime $\omega_-\to0^+$, the quadratic form
remains confining at every strictly positive $\omega_-$. Quadratic
confinement in the $(-)$ sector is lost only at the exact endpoint
$\omega_-=0$.

For a given value of $p_c(t)$, the instantaneous eigenstates of
$\hat H_1(t)$ in Eq.~(\ref{eq:H1}) take the form
\begin{equation}
    \ket{n_{\sss{+}}(t),n_{\sss{-}}(t)}
    =
    e^{\frac{i}{\hbar}p_0(t)(\hat x+\hat y)}\,
    e^{-\frac{i}{\hbar}x_0(t)(\hat p_x-\hat p_y)}\,
    \ket{n_{\sss{+}},n_{\sss{-}}}_{\scriptstyle 0}\,,
\end{equation}
as verified directly in Appendix~\ref{sec:S1}, up to a
state-dependent overall phase. Here
$\ket{n_{\sss{+}},n_{\sss{-}}}_{\scriptstyle 0}$ is an eigenstate of
$\hat H_0$ in Eq.~(\ref{eq:H0}), and the displacement parameters are
\begin{align}\label{eq:x0p0}
    x_0(t)
    &= \frac{\omega_2\,p_c(t)}
    {m(\omega_1^2-\omega_2^2)}\,,
    \qquad
    p_0(t)
    = -\frac{\omega_1^2\,p_c(t)}
    {\omega_1^2-\omega_2^2}\,.
\end{align}
These expressions are well defined for every $\omega_->0$ and finite
$p_c(t)$. The corresponding instantaneous eigenvalues are
\begin{subequations}\label{eq:eigen1}
\begin{align}
    E^{(1)}_{n_{\sss{+}},n_{\sss{-}}}(t)
    &=
    E^{(0)}_{n_{\sss{+}},n_{\sss{-}}}+c(t)
    \label{eq:eigen1a}
\end{align}
with
\begin{align}\label{eq:eigenenergy1}
    c(t)
    &=
    -\frac{\omega_2^2\,p_c(t)^2}
    {m(\omega_1^2-\omega_2^2)}\,.
\end{align}
\end{subequations}

Accordingly, the common shift $c(t)$ leaves all instantaneous level
differences unchanged. The normal-mode spacings remain
$\hbar\omega_+$ and $\hbar\omega_-$, respectively. At every strictly
positive $\omega_-$, the instantaneous Hamiltonian has a discrete
spectrum bounded below and admits a normalizable Gibbs state, as
shown in Sec.~\ref{sec:thermodynamic}. For fixed nonzero $p_c(t)$,
however, the displacement parameters $x_0(t)$ and $p_0(t)$ diverge
in magnitude and $c(t)\to-\infty$ as $\omega_1\to\omega_2>0$ from
above.

The shrinking soft-mode spacing $\hbar\omega_-$ also makes the
standard adiabatic criterion increasingly restrictive at a fixed
nonzero driving rate. Section~\ref{sec:nonadiabatic} quantifies this
drive- and occupation-dependent restriction within the finite-gap
regime. At the exact endpoint $\omega_-=0$, the loss of quadratic
confinement in one canonical sector requires a separate analysis of
the canonical trace, given in Sec.~\ref{sec:thermodynamic}.

\section{Nonadiabatic Amplification}\label{sec:nonadiabatic}
%
Spectral softening in the $(-)$ sector makes the standard adiabatic
criterion increasingly restrictive for the specified momentum-shift
drive. As derived in Appendix~\ref{sec:S2}, the exact nonadiabatic
couplings generated by $p_c(t)$ connect only adjacent levels in the
$(+)$ and $(-)$ normal modes, giving four possible off-diagonal
channels. For fixed occupations $n_{\sss{+}}$ and $n_{\sss{-}}$, the
$(-)$-mode raising channel imposes the most restrictive condition in
the quasi-soft limit:
\begin{equation}\label{eq:adiabacity condi1}
    \epsilon_{n_{\sss{-}}}(t)
    \equiv
    \frac{|\dot p_c(t)|}{\omega_{\sss{-}}^2}
    \sqrt{\frac{\omega_1}{2m\hbar}}
    \sqrt{n_{\sss{-}}+1}
    \ll 1\,,
\end{equation}
as given in Eq.~\eqref{eq:S2_adiabatic_mm}. At a given gap and
driving rate, $\epsilon_{n_{\sss{-}}}(t)$ increases with
$n_{\sss{-}}$. The least restrictive condition therefore occurs in
the soft-mode ground state, $n_{\sss{-}}=0$:
\begin{equation}
    \epsilon_0(t)
    =
    \frac{|\dot p_c(t)|}{\omega_{\sss{-}}^2}
    \sqrt{\frac{\omega_1}{2m\hbar}}
    \ll 1\,.
    \label{eq:ground_soft_condition}
\end{equation}
If the ground-state condition is violated, the corresponding
condition is also violated at every higher soft-mode occupation for
the same gap and driving rate.

To express the criterion in terms of frequency scales, we define
\begin{equation}
    \omega_{\sss{-}}^{c}(t)
    \equiv
    |\dot p_c(t)|^{1/2}
    \left(\frac{\omega_1}{2m\hbar}\right)^{1/4}\,.
    \label{eq:critical_soft_frequency}
\end{equation}
Since $\epsilon_0(t) =
[\omega_{\sss{-}}^{c}(t)/\omega_{\sss{-}}]^2$, the soft-mode
ground-state condition becomes
\begin{equation}
    \omega_{\sss{-}} \gg \omega_{\sss{-}}^{c}(t)\,.
    \label{eq:threshold_condition}
\end{equation}
In the regime
\begin{equation}
    0 < \omega_{\sss{-}}
    \lesssim \omega_{\sss{-}}^{c}(t)\,,
    \label{eq:breakdown_region}
\end{equation}
$\epsilon_0(t)$ is of order unity or larger, and the small-parameter
condition is no longer satisfied. Thus, for a nonzero driving rate,
the criterion can fail at a positive gap even though the Hamiltonian
remains bounded below, its spectrum remains discrete, and its Gibbs
state remains normalizable.

For an arbitrary soft-mode occupation, the same frequency scale
gives
\[
    \epsilon_{n_{\sss{-}}}(t)
    =
    \sqrt{n_{\sss{-}}+1}
    \left[
    \frac{\omega_{\sss{-}}^{c}(t)}{\omega_{\sss{-}}}
    \right]^2
    \ll 1\,.
\]
Equivalently, the condition is
\[
    \omega_{\sss{-}}
    \gg
    (n_{\sss{-}}+1)^{1/4}\omega_{\sss{-}}^{c}(t)\,.
\]
This reduces to Eq.~\eqref{eq:threshold_condition} for
$n_{\sss{-}}=0$. Higher soft-mode occupations, including those
populated thermally, impose a more restrictive condition.

The ground-state criterion can also be expressed as a separation of
timescales. For a nonzero instantaneous driving rate, define
\begin{equation}
    \tau_{\mathrm{int}}
    \equiv \frac{1}{\omega_{\sss{-}}}\,,
    \qquad
    \tau_{\mathrm{drive}}(t)
    \equiv \frac{1}{\omega_{\sss{-}}^{c}(t)}\,.
\end{equation}
Here $\tau_{\mathrm{drive}}(t)$ is a local driving timescale set by
$|\dot p_c(t)|$. The soft-mode ground-state parameter then satisfies
\begin{equation}
    \epsilon_0(t)
    =
    \left[
    \frac{\tau_{\mathrm{int}}}{\tau_{\mathrm{drive}}(t)}
    \right]^2\,,
\end{equation}
so the criterion requires
$\tau_{\mathrm{drive}}(t)\gg\tau_{\mathrm{int}}$. As the gap
decreases at a fixed nonzero driving rate, the increasing intrinsic
timescale makes this separation harder to maintain.

This criterion identifies when the local adiabatic parameter ceases
to be small for a given occupation. Transition amplitudes, however,
depend on the full driving protocol, including its duration,
waveform, and accumulated phases. Contributions from different times
may reinforce or cancel one another. Determining the probabilities
of transitions between instantaneous eigenstates therefore requires
calculating the time evolution under the specified protocol.
Regardless of the resulting transition probabilities, a violated
local criterion no longer provides a basis for assuming adiabatic
following.

At any fixed positive gap and occupation, the local criterion can be
satisfied by reducing the driving rate. For fixed $\omega_1$ and
$n_{\sss{-}}$, Eq.~\eqref{eq:adiabacity condi1} shows that
maintaining the same small value of $\epsilon_{n_{\sss{-}}}(t)$ as
the gap closes requires the driving rate to decrease in proportion
to $\omega_{\sss{-}}^2$. Halving the soft-mode gap therefore
requires reducing the driving rate by a factor of four. At a fixed
nonzero driving rate, however, $\epsilon_{n_{\sss{-}}}(t)$
eventually ceases to be small as $\omega_{\sss{-}}\to0^+$.

For an open system, the corresponding question is whether
bath-induced relaxation can sustain tracking of the instantaneous
Gibbs state under the imposed driving. The next section examines
this dynamical question at positive gaps and analyzes the loss of
canonical normalizability at the exact soft-mode endpoint.

\section{Thermodynamic Implications}\label{sec:thermodynamic}
%
\subsection{Thermal Relaxation}\label{subsec:thermal1}
Sec.~\ref{sec:nonadiabatic} showed how a shrinking soft-mode spacing
makes the isolated-system adiabatic criterion more restrictive. In
an open system, thermal tracking also depends on bath-induced
relaxation. The relaxation rate depends on the system--bath
interaction and the bath spectrum and cannot be inferred from the
soft-mode spacing alone.

To characterize relaxation in the soft sector, we hold $p_c$ fixed
at a strictly positive $\omega_-$ and work in the displaced
representation. In this representation, the system Hamiltonian in
Eq.~\eqref{eq:H1} takes the form $\hat H_0+c$. We consider the
illustrative linear interaction
\begin{equation}
    \hat H_{\mathrm{SB}}
    =
    \lambda(\hat a_-+\hat a_-^\dagger)\otimes\hat B\,,
    \label{eq:linearSB}
\end{equation}
where $\lambda$ sets the coupling strength and $\hat B$ is a
Hermitian bath operator. We assume $\langle\hat B\rangle_\beta=0$,
so the bath average of the interaction produces no static system
term at first order in $\lambda$. The linear system operator
connects only adjacent soft-mode levels, so the direct transition
rates retained at order $\lambda^2$ have $\Delta n_-=\pm1$
\cite{BreuerPetruccione,WeissBook}. Successive adjacent-level
transitions can connect more widely separated occupations. Direct
transitions with $|\Delta n_-|>1$ would require a different coupling
operator or higher-order processes beyond the treatment used here.

At each fixed positive gap, we assume weak coupling and bath
correlations that decay rapidly on the relaxation timescale. The
terms mixing the $+\omega_-$ and $-\omega_-$ components oscillate at
their difference frequency, $2\omega_-$. The secular approximation
averages out these terms when $2\omega_-$ is much larger than the
relevant dissipative rates \cite{BreuerPetruccione,WeissBook}. In
the resulting thermal amplitude-damping model, the soft-mode
populations evolve independently of the off-diagonal coherences in
the energy basis. As derived in Appendix~\ref{sec:S3}, the mean
occupation obeys
\begin{equation}
    \frac{d\langle\hat n_-\rangle}{dt}
    =
    -\kappa(\omega_-)
    \left(\langle\hat n_-\rangle-\bar n_-\right)\,,
    \qquad
    \bar n_-=
    \left(e^{\beta\hbar\omega_-}-1\right)^{-1}\,,
    \label{eq:meanrelax}
\end{equation}
where $\hat n_-=\hat a_-^\dagger\hat a_-$, $\beta=(k_{\mathrm
B}T)^{-1}>0$, and the positive damping coefficient
$\kappa(\omega_-)$ depends on the system--bath coupling and bath
correlations. At fixed $p_c$, Eq.~\eqref{eq:meanrelax} therefore
implies that the deviation of the mean soft-mode occupation
$\langle\hat n_-\rangle$ from its thermal value $\bar n_-$ decays
exponentially. The relaxation time
$\tau_{\mathrm{occ}}=1/\kappa(\omega_-)$ is the time for the
magnitude of this deviation to fall to $1/e$ of its initial
magnitude.

For the mean soft-mode occupation to approach its Gibbs value $\bar
n_-$, the relaxation time $\tau_{\mathrm{occ}}$ must be much shorter
than the time for which $p_c$ is held fixed. The secular timescale
separation assumed above requires, in particular, that the net
occupation-relaxation rate satisfy $\kappa(\omega_-)\ll 2\omega_-$.
It follows that $\tau_{\mathrm{occ}}\gg(2\omega_-)^{-1}$. As
$\omega_-\to0^+$, $(2\omega_-)^{-1}$ grows without bound. If $p_c$
is held fixed for the same duration at each gap, these two
requirements become incompatible for sufficiently small positive
$\omega_-$, where the corresponding Gibbs state remains
normalizable.

When $p_c(t)$ varies at fixed mode frequencies and bath temperature,
the normalized Boltzmann weights in the instantaneous energy basis
remain unchanged. The instantaneous eigenstates are displaced, so
the Gibbs density operator changes in the laboratory frame, as shown
in Appendix~\ref{sec:S1}. On intervals over which $p_c(t)$ varies
only slightly, the relaxation time $\tau_{\mathrm{occ}}$ obtained
from Eq.~\eqref{eq:meanrelax} at fixed $p_c$ provides a local
estimate of how quickly the bath reduces deviations of the mean
occupation from its Gibbs value. For these deviations to decay
substantially before $p_c(t)$ changes appreciably, the driving
timescale $\tau_{p_c}$ must be much longer than
$\tau_{\mathrm{occ}}$. Together with the secular timescale
separation assumed above, this requires
\[
    (2\omega_-)^{-1} \ll \tau_{\mathrm{occ}} \ll \tau_{p_c}\,.
\]
As $\omega_-\to0^+$, the timescale $(2\omega_-)^{-1}$ grows without
bound. If the driving timescale $\tau_{p_c}$ is held constant as the
gap closes, this separation of timescales cannot be maintained for
sufficiently small positive $\omega_-$, where the instantaneous
Gibbs state remains normalizable.

\subsection{Partition Function Divergence}\label{subsec:thermal2}
The preceding analysis addressed the conditions for dynamical
tracking. We now examine the existence of instantaneous canonical
equilibrium by evaluating the partition function as the soft-mode
frequency approaches zero. For every $\omega_->0$ in the confining
regime and finite $p_c(t)$, the spectrum in Eq.~\eqref{eq:eigen1a}
gives
\begin{equation}
    Z_1(t)
    \equiv
    \operatorname{Tr}e^{-\beta\hat H_1(t)}
    =
    e^{-\beta c(t)}Z_0\,,
    \qquad
    Z_0=Z_{\sss{+}}Z_{\sss{-}}\,,
\end{equation}
where
\[
    Z_{\sss{\pm}}
    =
    \sum_{n_{\sss{\pm}}=0}^{\infty}
    e^{-\beta\hbar\omega_{\sss{\pm}}
    (n_{\sss{\pm}}+1/2)}\,.
\]
At any finite positive temperature, both normal-mode sums converge
when $\omega_{\sss{+}}>0$ and $\omega_{\sss{-}}>0$. Since $c(t)$ is
finite at every such gap for finite $p_c(t)$, the canonical trace
$Z_1(t)$ is finite and the instantaneous Gibbs state is
normalizable.

The common energy shift contributes to the Helmholtz free energy
according to
\begin{equation}
    F_1(t)
    \equiv -\beta^{-1}\ln Z_1(t)
    =
    F_0+c(t)\,,
    \qquad
    F_0=-\beta^{-1}\ln Z_0\,.
\end{equation}
The common energy shift $c(t)$ changes the partition function by the
factor $e^{-\beta c(t)}$ and adds $c(t)$ to the Helmholtz free
energy. The same factor cancels upon normalization, leaving the
Boltzmann weights in the instantaneous energy basis unchanged at
fixed mode frequencies and bath inverse temperature $\beta$.

As $\omega_-\to0^+$ at fixed $\beta>0$, with $\omega_+$ approaching
a finite positive value, $Z_+$ approaches a finite positive limit
and the soft-mode factor diverges according to
\begin{equation}
    Z_-
    =
    \left[2\sinh(\beta\hbar\omega_-/2)\right]^{-1}
    \sim
    (\beta\hbar\omega_-)^{-1}\,.
    \label{eq:Zsoft_exact}
\end{equation}
Consequently, $Z_0\to\infty$ and $F_0\to-\infty$. For fixed nonzero
$p_c$, the shift $c(t)\to-\infty$ further enhances the divergence of
$Z_1(t)$ through the factor $e^{-\beta c(t)}$ and lowers $F_1(t)$ by
the same energy shift. At every strictly positive $\omega_-$,
however, the soft-mode probabilities
\[
    P_{n_-}
    =
    \left(1-e^{-\beta\hbar\omega_-}\right)
    e^{-n_-\beta\hbar\omega_-}\,,
    \qquad n_-=0,1,2,\cdots\,,
\]
sum to unity. As the gap decreases, the distribution spreads over
increasingly many levels while remaining normalized.

By contrast, the exact endpoint $\omega_-=0$, where
$\omega_1=\omega_2\equiv\omega>0$, requires a separate analysis
because the finite-gap displacement becomes singular there. For
$p_c=0$, the energy becomes independent of $n_-$, so each level of
the remaining $(+)$ mode has infinite degeneracy. The Hamiltonian
remains bounded below, but its canonical trace diverges on the
infinite two-dimensional plane. For $p_c\ne0$, completing the
squares in the endpoint Hamiltonian leaves the linear term $-\omega
p_c(\hat x-\hat y)$ along a direction with no quadratic confinement,
as shown in Appendix~\ref{sec:S1}. The energy can decrease without
bound along this direction. In both cases, the endpoint admits no
normalizable canonical Gibbs state.

Critical slowing down near continuous phase transitions concerns
relaxation and can coexist with a normalizable equilibrium measure
\cite{HohenbergHalperin1977}. For example, in a Landau description,
a positive quartic term in the effective order-parameter free energy
can preserve stability when the quadratic coefficient vanishes
\cite{Goldenfeld1992}. Here, the Hamiltonian is exactly quadratic,
so the vanishing soft-mode frequency leaves no higher-order
restoring term in that sector.

At positive gaps, the instantaneous Gibbs state is normalizable, and
whether it can be attained or tracked depends on the specified
driving and dissipation. At the exact soft-mode endpoint, the loss
of confinement precludes a normalizable canonical Gibbs state.

\section{PHASE-SPACE STRUCTURE AND WIGNER
REPRESENTATION}\label{sec:wigner}
%
To elucidate the geometric origin of the endpoint singularity, we
examine the Hamiltonian phase-space structure and its Wigner
representation. For one canonical coordinate--momentum pair, the
Wigner function of a density operator $\hat\rho$ is defined as
\cite{Wigner}
\begin{equation}
    W_\rho(x,p)
    =
    \frac{1}{2\pi\hbar}
    \int d\xi\; e^{-ip\xi/\hbar}
    \left\langle x+\frac{\xi}{2}\right|
    \hat\rho
    \left|x-\frac{\xi}{2}\right\rangle\,.
\end{equation}
This definition applies to both pure and mixed states, including
thermal states. The two-mode Wigner function is obtained by applying
the same transform to both canonical pairs.

\begin{figure}[t]
    \centering
    \includegraphics[width=\linewidth]{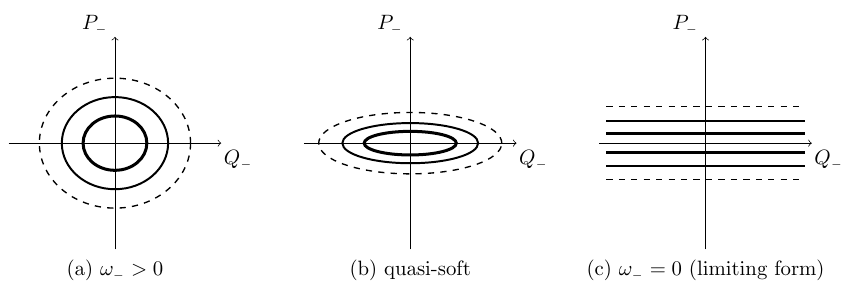}
    \caption{\justifying
    \textbf{Classical constant-energy contours of the $(-)$ normal
    mode for $p_c=0$.}
    The canonical coordinates $(Q_-,P_-)$ are defined in
    Eq.~\eqref{eq:S1_eigenstates0-0}. For $\omega_->0$, the mode
    Hamiltonian is $H_-=P_-^2/(2M)+M\omega_-^2Q_-^2/2$.
    (a) The contours are closed ellipses in the confining regime.
    (b) At fixed energy, the contours extend farther along $Q_-$
    as $\omega_-$ decreases and remain closed at every positive
    frequency.
    (c) In the formal $\omega_-\to0^+$ limit at fixed $(Q_-,P_-)$,
    the limiting form $H_-^{\mathrm{lim}}=P_-^2/(2M)$ yields
    horizontal contours. The coordinate transformation in
    Eq.~\eqref{eq:S1_eigenstates0-0} becomes singular at zero
    frequency. The exact endpoint of the original Hamiltonian
    is analyzed in Appendix~\ref{sec:S1}.}
    \label{fig:wigner_softmode}
\end{figure}

For $p_c=0$, Fig.~\ref{fig:wigner_softmode} illustrates the
weakening of quadratic confinement as $\omega_-$ decreases. This
weakening is also reflected in the thermal Wigner function derived
in Appendix~\ref{sec:S4}. In the confining regime, the function is
Gaussian, with widths determined by the temperature and the
normal-mode frequencies. At fixed inverse temperature $\beta>0$, the
width along the soft-mode coordinate $Q_-$ diverges as
$\omega_-\to0^+$. The distribution nevertheless remains normalized
at every strictly positive $\omega_-$, in agreement with the
canonical analysis in Sec.~\ref{subsec:thermal2}.

At fixed mode frequencies and temperature, varying $p_c(t)$
translates the instantaneous thermal Wigner function in phase space
while preserving its covariance and normalization. Thermal tracking
therefore requires the evolving state to follow the moving
equilibrium distribution even though the normalized Boltzmann
weights in the instantaneous energy basis remain unchanged.
Secs.~\ref{sec:nonadiabatic} and \ref{subsec:thermal1} address
complementary aspects of this problem through the isolated-system
response to driving and bath-induced relaxation and its implications
for slow driving, respectively.

In the classical limit $\hbar\to0$ at fixed positive gap, the
thermal Wigner function reduces to the normalized Boltzmann density
proportional to $e^{-\beta H_1}$, as shown in Appendix~\ref{sec:S4}.
This density undergoes the same phase-space translation as $p_c(t)$
varies, so the question of tracking a moving equilibrium
distribution also arises in the classical description. The classical
phase-space integral reproduces the equilibrium results of
Sec.~\ref{subsec:thermal2}. With the phase-space cell convention of
Appendix~\ref{sec:S4}, the classical partition function acquires the
factor $e^{-\beta c(t)}$, and the Helmholtz free energy shifts by
$c(t)$. At fixed temperature, the classical and quantum soft-mode
partition factors have the same leading $\omega_-^{-1}$ divergence
as $\omega_-\to0^+$. Both descriptions admit a normalized canonical
state at every positive gap. At the exact endpoint $\omega_-=0$, the
loss of confinement precludes a finite quantum canonical trace and a
finite classical canonical integral.

\section{Discussion}\label{sec:discussion}
%
Our analysis distinguishes the existence of instantaneous canonical
equilibrium from the dynamics required to attain or track it. For
every strictly positive $\omega_-$ and finite $p_c(t)$, the
instantaneous Hamiltonian has a discrete spectrum bounded below. At
every finite positive temperature, its canonical partition function
is finite and its Gibbs state is normalizable. These equilibrium
results hold, however small the positive gap becomes, independently
of the approximations used to describe thermal relaxation.

The exact endpoint $\omega_-=0$, where $\omega_1=\omega_2>0$, is
approached continuously from the confining regime. Along this
approach, the mean thermal occupation and phase-space width of the
soft mode grow, while the Gibbs state remains normalizable at every
positive gap. At the endpoint, quadratic confinement disappears in
one canonical sector. For $p_c=0$, the Hamiltonian remains bounded
below, with infinite degeneracy at each energy level. For $p_c\ne0$,
a residual linear potential along an unconfined direction makes the
Hamiltonian unbounded below. In both cases, the canonical trace on
the infinite plane fails to be finite.

Sec.~\ref{sec:nonadiabatic} treats unitary evolution in a thermally
isolated system. The prescribed drive can perform work on the
system, which exchanges no heat with its surroundings. Adiabatic
following here means that a state prepared in an instantaneous
energy eigenstate remains close, up to a phase, to the corresponding
instantaneous eigenstate as the drive varies. For the soft-mode
raising channel, the exact nonadiabatic coupling gives a local
adiabatic parameter proportional to $|\dot
p_c|\sqrt{n_-+1}/\omega_-^2$. At fixed occupation and fixed nonzero
driving rate, this parameter grows without bound as
$\omega_-\to0^+$, and the standard small-parameter criterion
eventually ceases to be satisfied. At each fixed positive gap,
sufficiently slow driving can satisfy the criterion. The resulting
restriction on the standard adiabatic approximation depends on both
the driving rate and the soft-mode occupation.

Sec.~\ref{sec:thermodynamic} considers an open system that can
exchange heat with a thermal environment. The illustrative linear
coupling permits direct transitions between adjacent soft-mode
levels at leading order. At fixed $p_c$, the weak-coupling,
Markovian, and secular treatment in Appendix~\ref{sec:S3} gives the
mean-occupation relaxation time
$\tau_{\mathrm{occ}}=1/\kappa(\omega_-)$. During quasistatic driving
at fixed mode frequencies and bath temperature, the changing
$p_c(t)$ displaces the instantaneous Gibbs state in phase space
while leaving its normalized Boltzmann weights unchanged. For the
bath to substantially reduce deviations of the mean occupation from
its Gibbs value before $p_c(t)$ changes appreciably, the driving
timescale must be much longer than $\tau_{\mathrm{occ}}$. Under the
secular timescale separation assumed in Sec.~\ref{subsec:thermal1},
this condition cannot be maintained if the driving timescale is held
constant as $\omega_-\to0^+$, even though the instantaneous Gibbs
state remains normalizable at every positive gap.

The thermal Wigner function and its classical limit give a common
phase-space interpretation of the equilibrium singularity. In both
descriptions, the thermal distribution broadens along the soft-mode
coordinate as the gap decreases and remains normalized at every
positive gap. At the exact endpoint, the loss of confinement
prevents a finite quantum canonical trace and a finite classical
canonical integral.

The same distinction is relevant to canonical work fluctuation
relations
\cite{Jarzynski1997,TalknerLutzHanggi2007,CampisiTalknerHanggi2011}.
Their standard formulations use normalized canonical preparations
and finite partition functions for the reference free energies, and
apply to driving away from the quasistatic regime. In the present
model, these equilibrium quantities are well defined at every
positive gap. This makes it possible to ask how work fluctuations
behave as the gap tends to zero from above. Protocols with the exact
soft-mode endpoint as a reference require separate treatment because
a normalizable canonical state is absent there.

This minimal quadratic model exhibits a loss of canonical
equilibrium through the collapse of a single normal-mode frequency.
At positive gaps, the driving protocol controls the conditions for
isolated-system adiabatic following, and quasistatic thermal
tracking also depends on bath-induced relaxation. At the exact
soft-mode endpoint, reversible quasistatic evolution through
canonical equilibrium states is precluded by the loss of equilibrium
itself.

\begin{acknowledgments}
The author gratefully acknowledges valuable discussions on
thermodynamic universality with G. J. Iafrate (North Carolina State
University) and M. von Spakovsky (Virginia Tech).
\end{acknowledgments}

\section*{Conflict of interest}
The author declares that there is no conflict of interest.

\section*{Data availability}
No datasets were generated or analyzed during the current study. All
results are analytical and can be reproduced from the equations and
methods presented in the manuscript.

\section*{References}
\bibliographystyle{apsrev4-2}
\bibliography{references}

\appendix


\section{Hamiltonian Structure and Instantaneous Eigenstates}
\label{sec:S1}
%
For $p_c=0$, we introduce the Cartesian ladder operators
\begin{equation}\label{eq:S1_ladders}
    \hat a_x
    =
    \sqrt{\frac{m\omega_1}{2\hbar}}\,\hat x
    +\frac{i\hat p_x}{\sqrt{2m\hbar\omega_1}}\,,
    \qquad
    \hat a_y
    =
    \sqrt{\frac{m\omega_1}{2\hbar}}\,\hat y
    +\frac{i\hat p_y}{\sqrt{2m\hbar\omega_1}}\,.
\end{equation}
The circular operators $\hat a_\pm=(\hat a_x\mp i\hat a_y)/\sqrt{2}$
obey $[\hat a_\sigma,\hat
a_{\sigma'}^\dagger]=\delta_{\sigma\sigma'}$. Writing $\hat
n_\pm=\hat a_\pm^\dagger\hat a_\pm$, we obtain $\hat L_z=\hbar(\hat
n_+-\hat n_-)$ and the diagonal Hamiltonian in Eq.~\eqref{eq:H0},
with $\omega_\pm=\omega_1\pm\omega_2$. The corresponding eigenstates
are
\begin{equation}\label{eq:S1_numberstates}
    \ket{n_+,n_-}_0
    =
    \frac{(\hat a_+^\dagger)^{n_+}(\hat a_-^\dagger)^{n_-}}
         {\sqrt{n_+!\,n_-!}}\ket{0,0}_0\,.
\end{equation}

To describe the normal modes in phase space, we first introduce
canonical variables scaled by the common frequency $\omega_1$,
\[
    \hat{\widetilde Q}_\pm
    =
    \sqrt{\frac{\hbar}{2M\omega_1}}\,
    (\hat a_\pm+\hat a_\pm^\dagger)\,,
    \qquad
    \hat{\widetilde P}_\pm
    =
    -i\sqrt{\frac{M\hbar\omega_1}{2}}\,
    (\hat a_\pm-\hat a_\pm^\dagger)\,,
    \qquad M=m/2\,.
\]
These variables remain well defined at $\omega_1=\omega_2>0$. In
these coordinates, each mode contributes
\[
    \hat H_{0,\pm}
    =
    \frac{\omega_\pm}{\omega_1}
    \left(
        \frac{\hat{\widetilde P}_\pm^2}{2M}
        +\frac{M\omega_1^2}{2}\hat{\widetilde Q}_\pm^2
    \right)\,.
\]
Thus the entire $(-)$-mode contribution to $\hat H_0$ vanishes at
the exact endpoint $\omega_-=0$.

In the confining regime $\omega_1>\omega_2>0$, we may also use the
frequency-scaled canonical variables
\begin{equation}\label{eq:S1_eigenstates0-0}
    \hat Q_\pm
    =
    \sqrt{\frac{\hbar}{2M\omega_\pm}}\,
    (\hat a_\pm+\hat a_\pm^\dagger)\,,
    \qquad
    \hat P_\pm
    =
    -i\sqrt{\frac{M\hbar\omega_\pm}{2}}\,
    (\hat a_\pm-\hat a_\pm^\dagger)\,.
\end{equation}
They satisfy $[\hat Q_\sigma,\hat
P_{\sigma'}]=i\hbar\delta_{\sigma\sigma'}$ and give $\hat
H_{0,\pm}=\hat P_\pm^2/(2M)+M\omega_\pm^2\hat Q_\pm^2/2$. The
eigenfunctions in the $(Q_+,Q_-)$ representation factorize into the
corresponding one-dimensional oscillator eigenfunctions. The
rescaling $Q_-=\sqrt{\omega_1/\omega_-}\,\widetilde Q_-$ and
$P_-=\sqrt{\omega_-/\omega_1}\,\widetilde P_-$ becomes singular at
$\omega_-=0$. Consequently, the form $P_-^2/(2M)$ illustrated in
Fig.~\ref{fig:wigner_softmode}(c) is the limit taken at fixed
$(Q_-,P_-)$. The exact endpoint of the original Hamiltonian is
obtained in the regular variables $(\widetilde Q_-,\widetilde P_-)$.

For finite $p_c(t)$ and $\omega_->0$, define the unitary
displacement
\begin{equation}\label{eq:S1_displacement}
    \hat D(t)
    =
    \exp\!\left[\frac{i}{\hbar}p_0(t)(\hat x+\hat y)\right]
    \exp\!\left[-\frac{i}{\hbar}x_0(t)(\hat p_x-\hat p_y)\right]\,.
\end{equation}
The two exponents commute because $[\hat x+\hat y,\hat p_x-\hat
p_y]=0$. Conjugation gives
\begin{equation}\label{eq:S1_shifts}
    \hat D^\dagger\hat x\hat D=\hat x+x_0\,,\quad
    \hat D^\dagger\hat y\hat D=\hat y-x_0\,,\quad
    \hat D^\dagger\hat p_x\hat D=\hat p_x+p_0\,,\quad
    \hat D^\dagger\hat p_y\hat D=\hat p_y+p_0\,.
\end{equation}
Direct substitution into Eq.~\eqref{eq:H1} yields
\begin{equation}\label{eq:S1_transformed}
    \hat D^\dagger\hat H_1\hat D
    = \hat H_0
    +\left[\frac{p_0+p_c}{m}+\omega_2x_0\right]
    (\hat p_x+\hat p_y)
    + \left[m\omega_1^2x_0+\omega_2p_0\right]
    (\hat x-\hat y)+C\,,
\end{equation}
where
\begin{equation}\label{eq:S1_C}
    C
    =
    \frac{(p_0+p_c)^2}{m}
    +m\omega_1^2x_0^2+2\omega_2x_0p_0\,.
\end{equation}
The linear terms vanish when
\begin{equation}\label{eq:S1_cancel}
    p_0+p_c=-m\omega_2x_0\,,
    \qquad
    m\omega_1^2x_0=-\omega_2p_0\,.
\end{equation}
Solving these equations gives the displacement parameters in
Eq.~\eqref{eq:x0p0} and the exact identity
\begin{equation}\label{eq:S1_unitaryidentity}
    \hat D^\dagger(t)\hat H_1(t)\hat D(t)=\hat H_0+c(t)\,,
\end{equation}
with $c(t)$ given by Eq.~\eqref{eq:eigenenergy1}. It follows that
$\hat D(t)\ket{n_+,n_-}_0$ are instantaneous eigenstates with
eigenvalues given by Eq.~\eqref{eq:eigen1a}. The same displacement
places the minimum of the classical Hamiltonian at
$(x,y,p_x,p_y)=(x_0,-x_0,p_0,p_0)$, where the Hamiltonian takes the
value $c(t)$.

At fixed mode frequencies and inverse temperature $\beta>0$,
Eq.~\eqref{eq:S1_unitaryidentity} gives $Z_1(t)=e^{-\beta c(t)}Z_0$,
where $Z_0=\operatorname{Tr}e^{-\beta\hat H_0}$. The normalized
instantaneous Gibbs density operator in the laboratory frame is
\[
    \hat\rho_{\beta,1}(t)
    =
    \frac{e^{-\beta\hat H_1(t)}}{Z_1(t)}
    =
    \hat D(t)\,
    \frac{e^{-\beta\hat H_0}}{Z_0}\,
    \hat D^\dagger(t)\,.
\]
The common energy shift $c(t)$ cancels upon normalization. The Gibbs
density operator is therefore time independent in the displaced
representation, while $\hat D(t)$ determines how it moves in the
laboratory frame as $p_c(t)$ varies. This displacement defines the
moving equilibrium state whose thermal tracking is discussed in
Sec.~\ref{subsec:thermal1}.

At the exact endpoint $\omega_1=\omega_2\equiv\omega>0$, consider
first $p_c=0$. The Hamiltonian is $\hat H_0=2\hbar\omega(\hat
n_++1/2)$, so each $(+)$-mode level has infinite degeneracy on the
infinite plane. The Hamiltonian remains bounded below, and the
factor $\operatorname{Tr}_-\hat I_-=\infty$ makes the canonical
trace diverge. For $p_c\ne0$ at the same endpoint, the cancellation
equations in Eq.~\eqref{eq:S1_cancel} require both $p_0+p_c=-m\omega
x_0$ and $p_0=-m\omega x_0$. These equations are compatible only for
$p_c=0$, so the finite-gap displacement construction fails.
Completing the squares in the original Hamiltonian gives the exact
identity
\begin{equation}\label{eq:S1_endpoint}
    \hat H_1
    =
    \frac{(\hat p_x+p_c-m\omega\hat y)^2}{2m}
    +\frac{(\hat p_y+p_c+m\omega\hat x)^2}{2m}
    -\omega p_c(\hat x-\hat y)\,.
\end{equation}
Expanding the two squares produces the cross terms $-\omega p_c\hat
y$ and $+\omega p_c\hat x$. The residual linear term cancels their
sum, recovering the original Hamiltonian.

To examine the energy along the unconfined direction, hold $p_c$
fixed and let $s\in\mathbb R$ be a displacement with units of
length. Define the Weyl phase-space translation \cite{Hillery} by
setting $x_0=s$ and $p_0=-m\omega s$ in the operator form of
Eq.~\eqref{eq:S1_displacement},
\[
    \hat T_s
    =
    \exp\!\left[-\frac{i}{\hbar}m\omega s(\hat x+\hat y)\right]
    \exp\!\left[-\frac{i}{\hbar}s(\hat p_x-\hat p_y)\right]\,.
\]
With $\hat D$ replaced by $\hat T_s$ in Eq.~\eqref{eq:S1_shifts},
the map $\hat O\mapsto\hat T_s^\dagger\hat O\hat T_s$ shifts $(\hat
x,\hat y)$ to $(\hat x+s,\hat y-s)$ and shifts both momenta by
$-m\omega s$. The operators inside the squares in
Eq.~\eqref{eq:S1_endpoint} then transform as
\begin{align*}
    (\hat p_x-m\omega s)+p_c-m\omega(\hat y-s)
    &=\hat p_x+p_c-m\omega\hat y\,,\\
    (\hat p_y-m\omega s)+p_c+m\omega(\hat x+s)
    &=\hat p_y+p_c+m\omega\hat x\,.
\end{align*}
Both squared terms are therefore invariant. Since $\hat x-\hat y$
shifts to $\hat x-\hat y+2s$, the residual linear term gives
\[
    \hat T_s^\dagger\hat H_1\hat T_s
    =
    \hat H_1-2\omega p_cs\,.
\]
For any normalized state $\ket{\psi}$ with a smooth wave function of
compact support, the energy expectation in the translated state
$\hat T_s\ket{\psi}$ is $\langle\psi|\hat H_1|\psi\rangle-2\omega
p_cs$. Taking $s\to+\infty$ for $p_c>0$ or $s\to-\infty$ for $p_c<0$
drives this expectation to $-\infty$. The endpoint Hamiltonian with
$p_c\ne0$ is therefore unbounded below.

Thus, at $\omega_-=0$, infinite degeneracy precludes a normalizable
canonical Gibbs state for $p_c=0$, while the unbounded linear
direction precludes it for $p_c\ne0$. For every strictly positive
$\omega_-$, Eq.~\eqref{eq:S1_unitaryidentity} yields a Hamiltonian
bounded below and a finite canonical partition function for
$\beta>0$. For fixed nonzero $p_c$, the spectral lower bound is
finite at each positive gap and tends to $-\infty$ as
$\omega_-\to0^+$.

\section{Exact Nonadiabatic Couplings and the Local Adiabatic Criterion}
\label{sec:S2}
%
We work in the confining regime $\omega_1>\omega_2>0$ and hold
$\omega_1$ and $\omega_2$ fixed during each driving protocol. Let
$\mathbf n=(n_+,n_-)$ and $\ket{\mathbf n(t)}=\hat D(t)\ket{\mathbf
n}_0$. Differentiating the instantaneous eigenvalue equation $\hat
H_1(t)\ket{\mathbf n(t)}=E_{\mathbf n}(t)\ket{\mathbf n(t)}$ and
projecting onto $\langle\mathbf m(t)|$ gives the standard identity
\cite{Messiah1962}
\begin{equation}\label{eq:S2_derivative}
 \langle\mathbf m(t)|\dot{\mathbf n}(t)\rangle
 =
 \frac{\langle\mathbf m(t)|\dot{\hat H}_1(t)|\mathbf n(t)\rangle}
 {E_{\mathbf n}(t)-E_{\mathbf m}(t)}\,,
 \qquad E_{\mathbf n}(t)\ne E_{\mathbf m}(t)\,.
\end{equation}
The condition on the energies is essential because distinct
occupation labels can have the same energy. The directly coupled
pairs identified below have an absolute energy difference of
$\hbar\omega_+$ or $\hbar\omega_-$.

The Hamiltonian derivative is
\begin{equation}\label{eq:S2_Hdot}
 \dot{\hat H}_1(t)
 =
 \frac{\dot p_c(t)}{m}
 [\hat p_x+\hat p_y+2p_c(t)]\,.
\end{equation}
Using the displacement in Appendix~\ref{sec:S1}, we obtain $\hat
D^\dagger\dot{\hat H}_1\hat D =\dot p_c[\hat p_x+\hat
p_y+2(p_0+p_c)]/m$. Only $\hat p_x+\hat p_y$ contributes to
off-diagonal matrix elements. In the canonical coordinates of
Eq.~\eqref{eq:S1_eigenstates0-0}, with $M=m/2$,
\begin{equation}\label{eq:S2_momentum}
 \hat p_x+\hat p_y
 =
 \sqrt{\frac{\omega_1}{\omega_+}}\hat P_+
 +\sqrt{\frac{\omega_1}{\omega_-}}\hat P_-
 +\frac{m}{2}\sqrt{\omega_1\omega_+}\hat Q_+
 -\frac{m}{2}\sqrt{\omega_1\omega_-}\hat Q_-\,.
\end{equation}
Each term is linear in the ladder operators of a single mode, giving
four possible off-diagonal channels.
\begin{subequations}\label{eq:S2_channels}
\begin{align}
 (n_+,n_-)&\longrightarrow(n_++1,n_-)\,,\quad(n_+-1,n_-)\,,\\
 (n_+,n_-)&\longrightarrow(n_+,n_-+1)\,,\quad(n_+,n_--1)\,.
\end{align}
\end{subequations}
A lowering channel is absent when the initial occupation of that
mode is zero. Define $\mathcal M_{\mathbf m\mathbf n}(t)
=\langle\mathbf m(t)|\dot{\hat H}_1(t)|\mathbf n(t)\rangle$.
Substituting Eq.~\eqref{eq:S2_momentum} into the off-diagonal part
$\dot p_c(\hat p_x+\hat p_y)/m$ and applying the ladder operators to
the initial number state gives
\begin{subequations}\label{eq:S2_elements}
\begin{align}
 \mathcal M_{(n_++1,n_-),(n_+,n_-)}
 &=\frac{\dot p_c}{2}\sqrt{\frac{\hbar\omega_1}{m}}
 (1+i)\sqrt{n_++1}\,,\\
 \mathcal M_{(n_+-1,n_-),(n_+,n_-)}
 &=\frac{\dot p_c}{2}\sqrt{\frac{\hbar\omega_1}{m}}
 (1-i)\sqrt{n_+}\,,\\
 \mathcal M_{(n_+,n_-+1),(n_+,n_-)}
 &=-\frac{\dot p_c}{2}\sqrt{\frac{\hbar\omega_1}{m}}
 (1-i)\sqrt{n_-+1}\,,\\
 \mathcal M_{(n_+,n_--1),(n_+,n_-)}
 &=-\frac{\dot p_c}{2}\sqrt{\frac{\hbar\omega_1}{m}}
 (1+i)\sqrt{n_-}\,.
\end{align}
\end{subequations}
Taking the absolute values of these matrix elements, we obtain
\begin{subequations}\label{eq:S2_moduli}
\begin{align}
 |\mathcal M_{(n_+\pm1,n_-),(n_+,n_-)}|
 &=|\dot p_c|\sqrt{\frac{\hbar\omega_1}{2m}}
 \begin{cases}
  \sqrt{n_++1}&\text{raising}\,,\\
  \sqrt{n_+}&\text{lowering}\,,
 \end{cases}\\
 |\mathcal M_{(n_+,n_-\pm1),(n_+,n_-)}|
 &=|\dot p_c|\sqrt{\frac{\hbar\omega_1}{2m}}
 \begin{cases}
  \sqrt{n_-+1}&\text{raising}\,,\\
  \sqrt{n_-}&\text{lowering}\,.
 \end{cases}
\end{align}
\end{subequations}
The common energy shift cancels from the raising-channel spacings,
\begin{equation}\label{eq:S2_gaps}
 E_{n_++1,n_-}-E_{n_+,n_-}=\hbar\omega_+\,,\qquad
 E_{n_+,n_-+1}-E_{n_+,n_-}=\hbar\omega_-\,.
\end{equation}
The corresponding lowering channels have the same absolute spacings.

The standard local adiabatic criterion compares the magnitude of
each off-diagonal derivative coupling with the corresponding
absolute Bohr frequency. Using Eq.~\eqref{eq:S2_derivative}, this
condition takes the form
\begin{equation}\label{eq:S2_localcriterion}
 \frac{\hbar|\mathcal M_{\mathbf m\mathbf n}(t)|}
 {|E_{\mathbf n}(t)-E_{\mathbf m}(t)|^2}\ll1\,.
\end{equation}
Substituting the matrix-element moduli and the mode spacings gives
the raising and lowering conditions for each mode.
\begin{subequations}\label{eq:S2_inequalities}
\begin{align}
 \frac{|\dot p_c|}{\omega_+^2}
 \sqrt{\frac{\omega_1}{2m\hbar}}\sqrt{n_++1}
 &\ll1\,,\qquad
 \frac{|\dot p_c|}{\omega_+^2}
 \sqrt{\frac{\omega_1}{2m\hbar}}\sqrt{n_+}\ll1\,,
 \label{eq:S2_adiabatic_pp}\\
 \frac{|\dot p_c|}{\omega_-^2}
 \sqrt{\frac{\omega_1}{2m\hbar}}\sqrt{n_-+1}
 &\ll1\,,\qquad
 \frac{|\dot p_c|}{\omega_-^2}
 \sqrt{\frac{\omega_1}{2m\hbar}}\sqrt{n_-}\ll1\,.
 \label{eq:S2_adiabatic_mm}
\end{align}
\end{subequations}
At fixed occupations, the $(-)$-mode raising channel gives the most
restrictive condition as $\omega_-\to0^+$ with $\omega_+$
approaching a finite positive value. This channel gives
Eq.~\eqref{eq:adiabacity condi1}. Away from this limit, a
sufficiently large $n_+$ can make the $(+)$-mode raising condition
more restrictive.

Writing $\epsilon_{n_-}(t)=\sqrt{n_-+1} [\omega_-^c(t)/\omega_-]^2$
gives the condition $\omega_-\gg(n_-+1)^{1/4}\omega_-^c(t)$. This
frequency scale marks where the local adiabatic parameter for a
given occupation becomes of order unity. Transition amplitudes
depend on the full driving protocol and the phases accumulated
during the evolution. A violation of
Eq.~\eqref{eq:S2_localcriterion} removes the basis for assuming
adiabatic following from this criterion alone. Determining the
transition probabilities requires the time evolution under the
specified protocol. At fixed positive gap and fixed occupation, a
sufficiently small $|\dot p_c|$ satisfies the local criterion. At
fixed nonzero $|\dot p_c|$ and fixed $n_-$, the soft-mode raising
parameter $\epsilon_{n_-}(t)$ grows without bound as
$\omega_-\to0^+$ with $\omega_1$ approaching a finite positive
value.

\section{Derivation of the Mean-Occupation Relaxation Equation}
\label{sec:S3}
%
We hold $p_c$ fixed and work in the displaced representation, where
the system Hamiltonian is $\hat H_0+c$. We use the interaction in
Eq.~\eqref{eq:linearSB} and the zero-mean bath condition stated
below that equation. Let $\ket{n}$ denote a number state of the
$(-)$ mode. The nonzero matrix elements of the system coupling
operator are
\begin{equation}\label{eq:S3_raise}
    \langle n+1|(\hat a_-+\hat a_-^\dagger)|n\rangle
    =\sqrt{n+1}\quad(n\geq0)\,,
    \qquad
    \langle n-1|(\hat a_-+\hat a_-^\dagger)|n\rangle
    =\sqrt n\quad(n\geq1)\,.
\end{equation}

For a stationary thermal bath, define the correlation spectrum
\[
    \gamma_\beta(\nu)
    =
    \frac{\lambda^2}{\hbar^2}
    \int_{-\infty}^{\infty}dt\,e^{i\nu t}
    \langle\hat B(t)\hat B(0)\rangle_\beta\,.
\]
Under the weak-coupling, Markovian, and secular approximations
specified in Sec.~\ref{subsec:thermal1}, the coefficients for
downward and upward transitions are
$\gamma_\downarrow=\gamma_\beta(\omega_-)$ and
$\gamma_\uparrow=\gamma_\beta(-\omega_-)$, respectively. A downward
transition $n\to n-1$ transfers energy $\hbar\omega_-$ from the mode
to the bath at rate $n\gamma_\downarrow$, and an upward transition
$n\to n+1$ transfers the same energy from the bath to the mode at
rate $(n+1)\gamma_\uparrow$. The thermal bath correlation spectrum
satisfies the detailed-balance relation
$\gamma_\beta(-\omega_-)=e^{-\beta\hbar\omega_-}
\gamma_\beta(\omega_-)$ \cite{BreuerPetruccione,WeissBook}. Thus
$\gamma_\uparrow=e^{-\beta\hbar\omega_-}\gamma_\downarrow$. For
nonzero bath spectral weight at $\omega_-$, define
\[
    \kappa(\omega_-)=\gamma_\downarrow-\gamma_\uparrow>0\,,
    \qquad
    \bar n_-=\frac{1}{e^{\beta\hbar\omega_-}-1}\,.
\]
Detailed balance then gives $\gamma_\uparrow=\kappa(\omega_-)\bar
n_-$.

Under the secular approximation, the soft-mode populations
$P_n=\langle n|\rho_-|n\rangle$, with $\rho_-$ the reduced state of
the $(-)$ mode, obey a closed rate equation. Level $n$ gains
population from $n+1$ at rate $(n+1)\gamma_\downarrow P_{n+1}$ and
from $n-1$ at rate $n\gamma_\uparrow P_{n-1}$. It loses population
to $n-1$ at rate $n\gamma_\downarrow P_n$ and to $n+1$ at rate
$(n+1)\gamma_\uparrow P_n$. Combining these contributions gives
\begin{equation}\label{eq:S3_population}
    \dot P_n
    = \gamma_\downarrow[(n+1)P_{n+1}-nP_n]
    + \gamma_\uparrow[nP_{n-1}-(n+1)P_n]\,,
    \qquad n\geq0\,,\quad P_{-1}\equiv0\,.
\end{equation}
Multiplying Eq.~\eqref{eq:S3_population} by $n$ and summing over
$n\geq0$ gives
\[
    \frac{d\langle\hat n_-\rangle}{dt}
    =
    -\gamma_\downarrow\langle\hat n_-\rangle
    +\gamma_\uparrow(\langle\hat n_-\rangle+1)\,.
\]
With $\kappa=\gamma_\downarrow-\gamma_\uparrow$ and
$\gamma_\uparrow=\kappa\bar n_-$, this reduces to
Eq.~\eqref{eq:meanrelax}.

As an explicit example, consider a bosonic reservoir with $\hat
H_B=\sum_k\hbar\nu_k\hat b_k^\dagger\hat b_k$ and $\hat
B=\sum_k(g_k\hat b_k+g_k^*\hat b_k^\dagger)$. In the continuum
limit, its positive-frequency spectral density is
$J_B(\nu)=\sum_k|g_k|^2\delta(\nu-\nu_k)$ for $\nu>0$. Within the
approximations used above, the thermal correlation spectrum gives
\[
    \kappa(\omega_-)
    =
    \gamma_\beta(\omega_-)-\gamma_\beta(-\omega_-)
    =
    \frac{2\pi\lambda^2}{\hbar^2}J_B(\omega_-)
\]
for the coupling in Eq.~\eqref{eq:linearSB}.

\section{Thermal Wigner Function and Quantum--Classical Correspondence}
\label{sec:S4}
%
For $\omega_->0$ and $\beta>0$, the normalized Gibbs state of $\hat
H_0$ is Gaussian. Let $\boldsymbol z=(x,y,p_x,p_y)^T$ and
$t_\pm\equiv\tanh(\beta\hbar\omega_\pm/2)$. The Wigner function is
the product of the two thermal oscillator Wigner functions
\cite{Hillery}, expressed here in the original Cartesian
coordinates:
\begin{equation}\label{eq:S4_Wigner}
    W_\beta^{(0)}(\boldsymbol z)
    =
    \frac{t_+t_-}{(\pi\hbar)^2}
    \exp\!\left[-\frac{2}{\hbar}
    \left(\frac{t_+}{\omega_+}H_+
         +\frac{t_-}{\omega_-}H_-\right)\right]\,.
\end{equation}
The classical mode Hamiltonians are
\begin{equation}\label{eq:S4_Hmodes}
    H_\pm
    =
    \frac{\omega_\pm}{4m\omega_1}(p_x^2+p_y^2)
    +\frac{m\omega_1\omega_\pm}{4}(x^2+y^2)
    \pm\frac{\omega_\pm}{2}(xp_y-yp_x)\,,
    \qquad H_++H_-=H_0\,.
\end{equation}
The normalization is $\int d^4z\,W_\beta^{(0)}=1$. For finite
$p_c(t)$, Eq.~\eqref{eq:S1_unitaryidentity} gives
$\hat\rho_{\beta,1}(t)= \hat D(t)\hat\rho_{\beta,0}\hat
D^\dagger(t)$. The coordinate and momentum shifts in
Eq.~\eqref{eq:S1_shifts} then determine the phase-space displacement
$\boldsymbol z_0(t)$ and give
\begin{equation}\label{eq:S4_translation}
    W_\beta^{(1)}(\boldsymbol z;t)
    =
    W_\beta^{(0)}(\boldsymbol z-\boldsymbol z_0(t))\,,
    \qquad
    \boldsymbol z_0(t)
    =
    (x_0(t),-x_0(t),p_0(t),p_0(t))^T\,.
\end{equation}
This translation preserves normalization and covariance at every
positive gap. The scalar energy shift cancels upon normalization,
while $Z_1(t)=e^{-\beta c(t)}Z_0$ and $F_1(t)=F_0+c(t)$.

As $\hbar\to0$ at fixed positive mode frequencies and temperature,
$t_\pm\sim\beta\hbar\omega_\pm/2$. Thus
\begin{equation}\label{eq:S4_classical0}
    W_\beta^{(0)}(\boldsymbol z)
    \longrightarrow
    W_{\beta,\mathrm{cl}}^{(0)}(\boldsymbol z)
    =
    \frac{e^{-\beta H_0(\boldsymbol z)}}{\mathcal I_\beta^{(0)}}\,,
    \qquad
    \mathcal I_\beta^{(0)}
    =
    \int d^4z\,e^{-\beta H_0(\boldsymbol z)}
    =
    \frac{(2\pi)^2}{\beta^2\omega_+\omega_-}\,.
\end{equation}
For $H_1(\boldsymbol z;t) =H_0(\boldsymbol z-\boldsymbol
z_0(t))+c(t)$, the factor $e^{-\beta c(t)}$ appears in both the
Boltzmann weight and its normalization integral and thus cancels in
the normalized density. Using Eq.~\eqref{eq:S4_translation}, we
obtain
\begin{equation}\label{eq:S4_classical1}
    W_\beta^{(1)}(\boldsymbol z;t)
    \longrightarrow
    W_{\beta,\mathrm{cl}}^{(1)}(\boldsymbol z;t)
    =
    \frac{e^{-\beta H_1(\boldsymbol z;t)}}{\mathcal I_\beta^{(1)}(t)}\,,
    \qquad
    \mathcal I_\beta^{(1)}(t)
    =
    e^{-\beta c(t)}\mathcal I_\beta^{(0)}\,.
\end{equation}
The phase-space integral $\mathcal I_\beta$ normalizes the classical
density with respect to $d^4z$. Dividing by the phase-space cell
volume $(2\pi\hbar)^2$ gives the dimensionless classical partition
function
\begin{equation}\label{eq:S4_Zclassical}
    Z_{\mathrm{cl}}^{(1)}(t)
    =
    \frac{\mathcal I_\beta^{(1)}(t)}{(2\pi\hbar)^2}
    =
    \frac{e^{-\beta c(t)}}{\beta^2\hbar^2\omega_+\omega_-}\,.
\end{equation}
For comparison, the exact quantum soft-mode factor is
\begin{equation}\label{eq:S4_Zsoft}
    Z_-
    =
    \sum_{n=0}^\infty e^{-\beta\hbar\omega_-(n+1/2)}
    =
    \frac{1}{2\sinh(\beta\hbar\omega_-/2)}
    \sim\frac{1}{\beta\hbar\omega_-}
    \qquad(\omega_-\to0^+)\,.
\end{equation}
It is finite at every positive $\omega_-$ and has the same leading
soft-mode scaling as the classical factor at fixed temperature.

The broadening is quantified by the Gaussian variances in the
normal-mode coordinates of Eq.~\eqref{eq:S1_eigenstates0-0},
\begin{equation}\label{eq:S4_normalvariances}
    \operatorname{Var}(Q_\pm)
    =
    \frac{\hbar}{2M\omega_\pm}
    \coth\!\frac{\beta\hbar\omega_\pm}{2}\,,
    \qquad
    \operatorname{Var}(P_\pm)
    =
    \frac{M\hbar\omega_\pm}{2}
    \coth\!\frac{\beta\hbar\omega_\pm}{2}\,.
\end{equation}
At fixed $\beta>0$, $\operatorname{Var}(Q_-) \sim1/(\beta
M\omega_-^2)$ and $\operatorname{Var}(P_-)\to M/\beta$ as
$\omega_-\to0^+$. Broadening is also present in the original
Cartesian coordinates, which remain well defined at the endpoint:
\begin{align}\label{eq:S4_cartesianvariances}
    \operatorname{Var}(x)=\operatorname{Var}(y)
    &=
    \frac{\hbar}{4m\omega_1}
    \left[
    \coth\!\frac{\beta\hbar\omega_+}{2}
    +\coth\!\frac{\beta\hbar\omega_-}{2}
    \right]\,,\nonumber\\
    \operatorname{Var}(p_x)=\operatorname{Var}(p_y)
    &=
    \frac{m\hbar\omega_1}{4}
    \left[
    \coth\!\frac{\beta\hbar\omega_+}{2}
    +\coth\!\frac{\beta\hbar\omega_-}{2}
    \right]\,.
\end{align}
Thus, as $\omega_1\to\omega_2>0$, the Cartesian position variance
grows as $1/(2\beta m\omega_1\omega_-)$, while the distribution
remains normalized at every positive gap.
Eq.~\eqref{eq:S4_translation} shifts the mean and preserves these
variances.

At the exact endpoint $\omega_-=0$, the Hamiltonian is analyzed
directly in Appendix~\ref{sec:S1}. For $p_c=0$, the infinite
degeneracy of the zero-frequency sector makes the quantum canonical
trace diverge, and the corresponding unconstrained classical sector
gives an infinite phase-space integral. For $p_c\ne0$, the residual
linear term allows the energy to decrease without bound, which also
precludes canonical normalization. In both cases, the loss of
confinement in one canonical sector prevents a normalized canonical
thermal distribution in both the quantum and classical descriptions.
\end{document}